\documentclass[aps,prc,twocolumn,groupedaddress]{revtex4-1}
\usepackage{graphicx}
\usepackage{amsmath}
\usepackage{braket}
\usepackage{url}
\usepackage{multirow}
\usepackage{multirow}
\usepackage{amssymb}
\usepackage{booktabs}
\usepackage{rotating}
\usepackage{bm}
\usepackage{bbm}

\usepackage[colorlinks,
linkcolor=blue,
anchorcolor=red,
urlcolor=red,
citecolor=red]{hyperref}

\begin{document}

\title{Investigation of unbound hydrogen isotopes with the Gamow shell model}

\author{H. H. Li}
\author{J. G. Li}
\author{N. Michel}\email[]{nicolas.michel@impcas.ac.cn}
\author{W. Zuo}
\affiliation{Institute of Modern Physics, Chinese Academy of Sciences, Lanzhou 730000, China}
\affiliation{School of Nuclear Science and Technology, University of Chinese Academy of Sciences, Beijing 100049, China}

\date{\today}

\begin{abstract}

Although they are part of the lightest nuclei, the hydrogen isotopes are not well understood both experimentally and theoretically. Indeed, besides deuteron and triton, all known hydrogen isotopes are resonances of complex structure. Even more elusive is $^7$H, which may have been observed experimentally and has been claimed to be a narrow resonance. Nevertheless, even its existence is controversial, and its theoretical study is difficult due to both its unbound character and large number of interacting valence nucleons. It is then the object of this paper to theoretically study the hydrogen isotopes $^{4-7}$H with the Gamow shell model, which is, up to our knowledge, the first direct calculation of unbound resonance hydrogen isotopes up to $^7$H. As the Gamow shell model includes both continuum coupling and inter-nucleon correlations, useful information can be obtained about poorly known unbound hydrogen isotopes. Our present calculations indicate that $^{4,6}$H ground states are fairly broad resonances, whereas those of $^{5,7}$H are narrow, which is in accordance with current experimental data. The results then suggest that, in particular, $^{5,7}$H should be more heavily studied, as they might well be among the most narrow neutron resonances of the light nuclear chart.

\end{abstract}


\maketitle

\textit{Introduction.}~  
The lightest nuclei are among the most interesting nucleonic systems of the nuclear chart \cite{PhysRevLett.116.052501,PhysRevLett.124.022502,PhysRevLett.108.052504}. The structure of the lightest nuclei is typically very complex \cite{LEIDEMANN2013158,PhysRevLett.119.032501,PhysRevC.100.054313,PhysRevC.104.024319,PhysRevC.88.044318}. A detailed description of the structure of $^{4,5}$He could be obtained with realistic interactions in a no-core framework \cite{PhysRevC.88.044318}. Several theoretical approaches have been devised to study light nuclei, where realistic interactions are included: Faddeev-Yakubovski equations \cite{PhysRevC.93.044004,10.3389/fphy.2019.00251,Lazauskas2019}, no-core shell model with harmonic oscillator \cite {BARRETT2013131,PhysRevC.95.014306,PhysRevC.86.034325,PhysRevC.79.014308} or Berggren bases \cite{PhysRevC.88.044318,PhysRevC.104.024319}, coupled-channel potential equations generated microscopically from no-core shell model \cite{PhysRevC.103.035801,PhysRevC.100.024304,PhysRevC.97.034332,PhysRevC.97.034314,PhysRevLett.114.212502}, complex scaling of realistic Hamiltonians \cite{PhysRevC.104.034612,PhysRevC.86.044002,PhysRevC.93.044004,PhysRevC.85.054002,PhysRevC.86.044002}, ... Added to that, as light nuclei can be accessed experimentally up to drip-lines and also beyond \cite{PhysRevLett.124.022502}, they form ideal laboratories to study the nucleon-nucleon interaction from both experimental and theoretical points of views.

However, the hydrogen chain, besides the deuteron and triton isotopes \cite{PhysRevC.48.792,PhysRevC.51.38,PhysRevC.63.024001,PhysRevC.66.014002,PhysRevLett.113.252001,PhysRevC.92.054002,PhysRevC.99.054003,PhysRevC.96.024001,PhysRevC.99.024313,PhysRevC.103.054001}, has been typically ignored in the vast amount of theoretical calculations devoted to light nuclei. Two main reasons can explain this situation. Firstly, hydrogen isotopes bear a single proton. Thus, it is impossible to represent a hydrogen wave function, even approximately, by a few valence nucleons interacting above a even-even core nucleus coupled to $J=0$. In fact, this core+valence nucleon picture, fundamental in shell model \cite{RevModPhys.77.427}, demands at least to use a $^4$He core for the lightest nuclei. Consequently, hydrogen isotopes should ideally be described in a no-core picture. Secondly, hydrogen isotopes are all unbound for $A \geq 4$, so that models devised for well bound nuclei, such as the harmonic oscillator no-core shell model, cannot be used, insofar as neutron-emission width may reach a few MeV (see Tab.~\ref{H_isotopes_experimental}). Phenomenological calculations have been effected using either harmonic oscillator shell model in restricted model spaces \cite{Hees1983ATI,PhysRevC.33.699} or cluster models \cite{PhysRevC.63.027001,PhysRevC.68.034303,ADAHCHOUR2008252,Grigorenko_EPA,Grigorenko_EPA_5H,PhysRevC.62.014312,PhysRevC.69.034336,PhysRevC.97.034605,Arickx_2008}. In fact, the only realistic calculations for the unbound hydrogen isotopes were done using Faddeev-Yakubovski equations \cite{PhysRevC.71.034004,Lazauskas2019}. However, even in this last case, besides calculations of the reaction cross sections of n+$^3$H \cite{PhysRevC.71.034004}, only the widths of the $^{4,5}$H ground states have been assessed from extrapolation methods \cite{Lazauskas2019}. Added to that, calculating hydrogen nuclear wave functions is out of reach in the no-core Gamow shell model (no-core GSM), due to the quickly increasing model space dimensions, on the one hand, and to the difficulty to obtain convergence for broad resonances, on the other hand \cite{PhysRevC.100.054313,PhysRevLett.119.032501}. In fact, no-core GSM can be applied only up to $A=4$ without truncations \cite{PhysRevC.100.054313,PhysRevC.104.024319}. Indeed, the broad resonance formed by the tetraneutron ground state \cite{PhysRevC.100.054313}, as well as the unbound ground states of $^4$H and $^4$Li \cite{PhysRevC.104.024319}, could be successfully described in this framework with realistic interactions. Note, nevertheless, that the properties of unbound $A=4$ nuclei are still debated \cite{PhysRevC.104.024319}. In particular, the value of the neutron-emission width of $^4$H and $^4$Li is not precisely known experimentally, as it varies by several MeVs from an experiment to another \cite{PhysRevC.104.024319}.

\begin{table}[!htb]
\caption{ {Energies (E) and widths ($\Gamma$) of hydrogen isotopes $^{4-7}$H obtained in various experiments (Exp.). Energies are given in MeV and widths in keV. }} \label{H_isotopes_experimental}
\setlength{\tabcolsep}{1mm}{\begin{tabular}{ccccccc}
\hline \hline
& E& $\Gamma$&Exp.&E& $\Gamma$&Exp.\\ \hline
$^{4}\rm{H}$&1.99$\pm$0.37&2850$\pm$300&\cite{SIDORCHUK200454}
&1.6$\pm$0.1&400$\pm$100&\cite{2009PPN....40..558G}\\
~&2.7$\pm$0.6&2300$\pm$600&\cite{THOENNESSEN201243}\\ \hline

$^{5}\rm{H}$&$\approx 1.8$&$\approx 1500$&\cite{PhysRev.173.949}
&$\approx 1.8$&$\approx 1200$&\cite{PhysRev.173.949}\\
~&1.7 $\pm$ 0.3&1900$\pm$400&\cite{PhysRevLett.87.092501}
&1.8 $\pm$ 0.1& $<$500&\cite{GOLOVKOV200370}\\
~&1.8 $\pm$ 0.1& $<$600&\cite{SIDORCHUK2003C229}
&1.8$\pm$0.2&1300$\pm$500&\cite{STEPANTSOV2004436}\\
~&$\approx 2 $&~&\cite{PhysRevLett.93.262501}
&1.7$\pm$0.3&$\approx2500$&\cite{10.1007/3-540-37642-9_88}\\
~&$\approx 2 $&$\approx 1300 $&\cite{PhysRevC.72.064612}
&2.4 $\pm$ 0.3&4800 $\pm$400&\cite{PhysRevC.95.014310}\\ \hline

$^{6}\rm{H} $&2.6$\pm$0.5&1500$\pm$300&\cite{BELOZYOROV1986352}
&2.7$\pm$0.4&1800$\pm$500&\cite{THOENNESSEN201243}\\ \hline

$^{7}\rm{H}$&$\sim$1.7&$<$2000&\cite{Korsheninnikov_2000}
&$\approx 3.0$&$<$1000&\cite{GOLOVKOV2004163} \\
~&$\sim$2& &\cite{PhysRevC.81.064606}
&1.8(5)&$<$300&\cite{PhysRevLett.124.022502}\\
~&2.2(5)& &\cite{PhysRevC.103.044313}\\

\hline \hline 
\end{tabular}}
\end{table}

It is then the object of this Letter to evaluate the energies and neutron-emission widths of the unbound hydrogen isotopes with GSM. For this, we will firstly describe the model used for that matter, by focusing on the theoretical assumptions made. In order to assess the theoretical errors induced, different Hamiltonians will be considered. While the two-body interaction will be fixed from a calculation of helium isotopes similar to that done in Refs.\cite{PhysRevC.84.051304,PhysRevC.96.054316}, the one-body part, generated by the used core, will have its parameters varied. We will then consider in more details the broad or narrow character of unbound hydrogen isotopes, as we will see that both these types of resonances are present in the evaluated unbound hydrogen isotopes. The important case of $^7$H will be emphasized, in particular. Then, conclusions will be made in relation with the current experimental situation.

\textit{Method.}~As one cannot include a large number of valence nucleons in a GSM calculation, it is necessary to freeze a few nucleons in a core other than $^4$He. Hence, we will consider the core+valence neutron picture using a core of $^3$H.
As the $^3$H ground state is coupled to $J^\pi = 1/2^+$, one always obtains two degenerate many-body ground states of same energy but different total angular momentum unless valence neutrons are coupled to $J=0$. While this is problematic in general, this deficiency is rather mild in our case considering the simplicity of our model. On the one hand, valence neutrons are always coupled to $J^\pi = 0^+$ in $^{5,7}$H ground states, so that one always has $J^\pi = 1/2^+$ for the total angular momentum of $^{5,7}$H ground states therein. On the other hand, the doublet of lowest energy states $1^-$ and $2^-$ in $^4$H is almost degenerate, as they differ by about 300 keV experimentally \cite{ensdf}, and by only 50 keV in our previous no-core Gamow shell calculation \cite{PhysRevC.104.024319}, while their widths has been estimated to be about 1 MeV. However, the situation is unclear concerning $^6$H. The angular momentum of the ground state of $^6$H is not known experimentally and supposed to be either $2^-$ or $1^+$ \cite{ensdf}. However, a total angular momentum of $1^+$ was found by using harmonic oscillator shell model \cite{PhysRevC.33.699}, whose reliability is doubtful for broad resonance states.  In fact, one can demand the angular momentum of the ground state of $^6$H to be equal to $2^-$, which is, in fact, the most consistent assumption one can make in the frame of the used model.

The use of a core+valence nucleon picture also prevents from devising a realistic Hamiltonian for the calculation of hydrogen nuclear wave functions. In principle, it is possible to derive an effective Hamiltonian in a truncated valence model space from an initial realistic nucleon-nucleon interaction, using, for example, in-medium similarity group renormalization \cite{PhysRevLett.113.142501} or the extended Kuo-Krenciglowa method \cite{TAKAYANAGI201191}. However, assuming that the latter methods can be utilized with cores coupled to $J \neq 0$, convergence problems are expected to occur in the numerical calculation of the effective interaction because the many-body wave functions of hydrogen isotopes are well spread in momentum space \cite{SUN2017227}. Consequently, it is necessary to utilize phenomenological interactions whose parameters are fitted on experimental data. Hence, for that matter, we will use in the following the Minnesota (MN) force \cite{THOMPSON197753}, and the Furutani-Horiuchi-Tamagaki (FHT) interaction \cite{FHT1}, which have been successfully used for the description of helium isotopes in the frame of GSM \cite{PhysRevC.84.051304,PhysRevC.96.054316}. As experimental data related to unbound hydrogen isotopes are scarce, on the one hand, and as one aims at making predictions about $^7$H in particular, on the other hand, one will use MN and FHT interactions whose parameters have been optimized using $^4$He as a core to reproduce neutron-rich helium isotopes. Only the parameters of the Woods-Saxon (WS) potential mimicking the $^3$H core will be modified afterwards. Thus, in order for results to be predictive, the WS central potential depth will vary in a given range, so that the variations of energy and widths of unbound hydrogen isotopes can be assessed.

As such, we will use the core+valence particle picture with the Berggren basis \cite{BERGGREN1968265} to generate the many-body space within GSM (see \cite{0954-3899-36-1-013101} for a review of GSM). GSM relies on the completeness of the Berggren basis in the complex momentum plane for a given partial wave of quantum numbers $\ell j$:
\begin{equation}
\sum_{n} \ket{n \ell j} \bra{n \ell j} + \int_{L_+}  \ket{k \ell j} \bra{k \ell j} \, dk = \mathbbm{1},
\label{Berggren_comp}  
\end{equation}
where the $L^+$ integral contains the continuous part of the Berggren basis (see Ref.\cite{0954-3899-36-1-013101} for definition and details), with $\ket{k \ell j}$ running over the scattering states belonging to the $L^+$ contour of complex linear momenta, and where the discrete part, built from $\ket{n \ell j}$ states, consists of bound states and of the resonance states inside the $L^+$ contour. The $L^+$ integral is efficiently discretized with the Gauss-Legendre quadrature, as convergence is obtained with 30-50 points \cite{0954-3899-36-1-013101}. The GSM many-body basis of Slater determinants is then generated by occupying the one-body states of the discretized Berggren basis for all considered partial waves.

As we consider the core+valence particle picture, the most natural framework is that of the cluster orbital shell model (COSM) \cite{PhysRevC.38.410}. In COSM, all valence particle coordinates are defined with respect to the center of mass of the core, so that the COSM formalism is translationally invariant and then no spurious center of mass motion can occur. The Pauli principle is taken into account by demanding the occupied states of the core to be orthogonal to the valence one-body states \cite{OCM}. The two Hamiltonians considered for the helium and hydrogen isotopes bear the same formal structure in the COSM framework and read: 
\begin{eqnarray}
H=\displaystyle\sum_{i=1}^{A_{val}} \left( \dfrac{{\bm{p}_{i}}^{2}}{2\mu} + U_{core}(r_i)
\right) + \displaystyle\sum_{i<j}^{A_{val}}   \left(V_{ij} + \dfrac{\bm{p}_{i}\cdot\bm{p}_{j} }{M_{core}} \right).
\label{Hamiltonian}
\end{eqnarray}
where $A_{val}$ is the number of valence nucleons, $\mu$ is the effective mass of the core+nucleon two-body system, $U_{core}(r)$ is the WS potential mimicking the considered core, $V$ is the MN or FHT interaction and the last part of the Hamiltonian is the recoil term, arising from the finite mass of the core $M_{core}$.

The GSM Hamiltonian takes the form of a complex symmetric matrix to diagonalize. As many-body resonance states are hidden among scattering eigenstates, the overlap method is used to identify resonance eigenstates \cite{0954-3899-36-1-013101}. Very large Hamiltonian matrix dimensions occur due to the numerous discretized scattering states. It is then necessary to impose model space truncations when the Berggren basis is used. In our calculation, the number of nucleons occupying scattering one-body states has been limited to three, which is customary to write as 3p-3h truncations. This doing, GSM matrix dimensions are tractable only for $^{4-6}$H, but not for $^7$H, as the latter possesses four valence neutrons above the $^3$H core. In order to have complete calculations for $^7$H, we will generate the GSM many-body basis from natural orbitals \cite{NaturalOrbitals}. Natural orbitals are the eigenstates of the reduced density matrix, so that nucleon occupation is concentrated on the lowest natural orbitals \cite{NaturalOrbitals}. Consequently, one typically obtains convergence using only 5-7 natural orbitals for a given partial wave. In the following, we will then generate natural orbitals using 3p-3h truncations in $^7$H. Calculations without truncations can then be realized for $^7$H by using a basis of natural orbitals, whereby the GSM Hamiltonian matrix dimension is strongly reduced. More theoretical details and recent applications of the mentioned numerical techniques in the GSM framework can be found in Refs.\cite{0954-3899-36-1-013101,PhysRevC.96.054316}.\\


\textit{Results.}~
We will firstly consider the two-body interaction for helium isotopes, hence with a $^4$He core. The FHT interaction was fitted in Ref.\cite{PhysRevC.96.054316} and analyzed using statistical tools within linear regression in order to quantify theoretical errors. It is not the case for the MN interaction, which is a phenomenological interaction whose parameters were tailored to the study of the charge radii of $^{6,8}$He in Ref.\cite{PhysRevC.84.051304}. Consequently, in order to assess how predictive the MN interaction is, two different MN interactions will be used, one with the initial parameters of Ref.\cite{PhysRevC.84.051304}, denoted as MN1, and another MN interaction, denoted as MN2, described in the following.  The GSM model spaces are indeed different according the used MN1, MN2 or FHT interactions. For the MN1 interaction, the one-body basis consists of the Berggren basis partial waves bearing $\ell \leq 2$ while, for the MN2 and FHT interactions, it consists of the same $psd$ partial waves, to which the $f$ partial waves represented by harmonic oscillator shells are added, following the same approach as in Ref.\cite{PhysRevC.96.054316}. The two different model spaces have been introduced in order to assess the dependence on the number of partial waves. As a result, it was found that the MN2 interaction parameters must be equal to 1.1 times those of the initial MN interaction of Ref.\cite{PhysRevC.84.051304} for an optimal reproduction of helium binding energies and widths.

\begin{figure}[!htb]
\includegraphics[width=1.00\columnwidth]{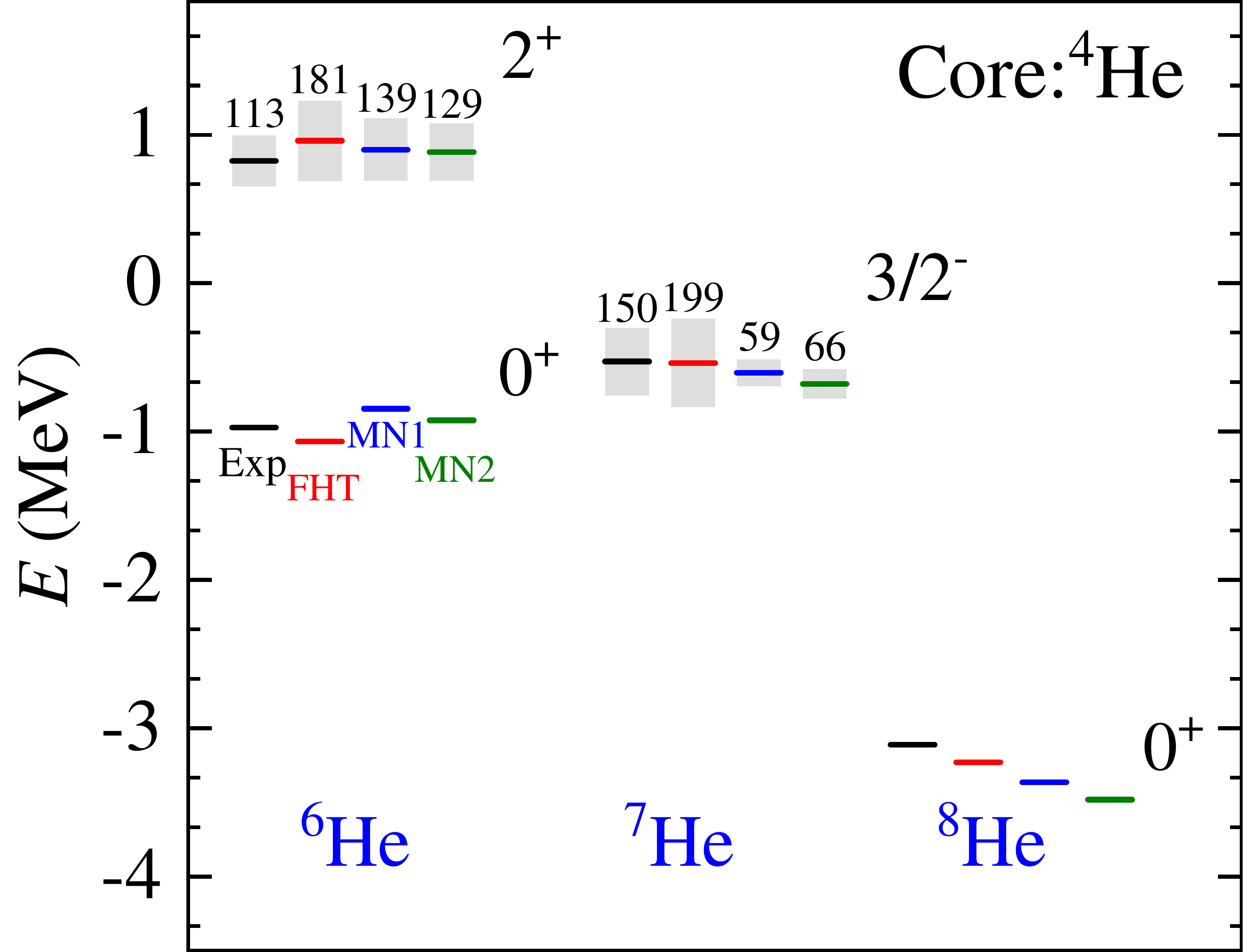}
\caption{Energy spectra and neutron-emission widths of helium isotopes $^{6-8}$He obtained with GSM using the FHT, MN1 and MN2 interactions (see text for definition) and compared to experimental data (Exp). Energies are in MeV. The numerical value of width is written above the energy in keV units. Width is also represented by a shaded area, enlarged by a factor three for clarity on screen. }{\label{He_FHT_MN1_MN2}}
\end{figure}

Results are shown in Fig.~\ref{He_FHT_MN1_MN2}, where one concentrates on $^{6-8}$He, as their number of valence neutrons is the same as in $^{5-7}$H, respectively. One can see, as could be expected, that the best reproduction of experimental data is obtained with the FHT interaction, where differences between the calculated and experimental energies and widths do not exceed 100 keV. While the energy and width of the $0^+$ and $2^+$ states of $^6$He are also well reproduced with MN interactions, the ground states of $^7$He and $^8$He are slightly too bound, with the width of $^7$He being about 90 keV too narrow. However, the results are overall satisfactory, as the largest discrepancy with experimental data occurs in $^8$He, where the binding energy obtained with MN2 is about 120 keV more bound than that obtained with MN1.

\begin{figure}[!htb]
\includegraphics[width=1.00\columnwidth]{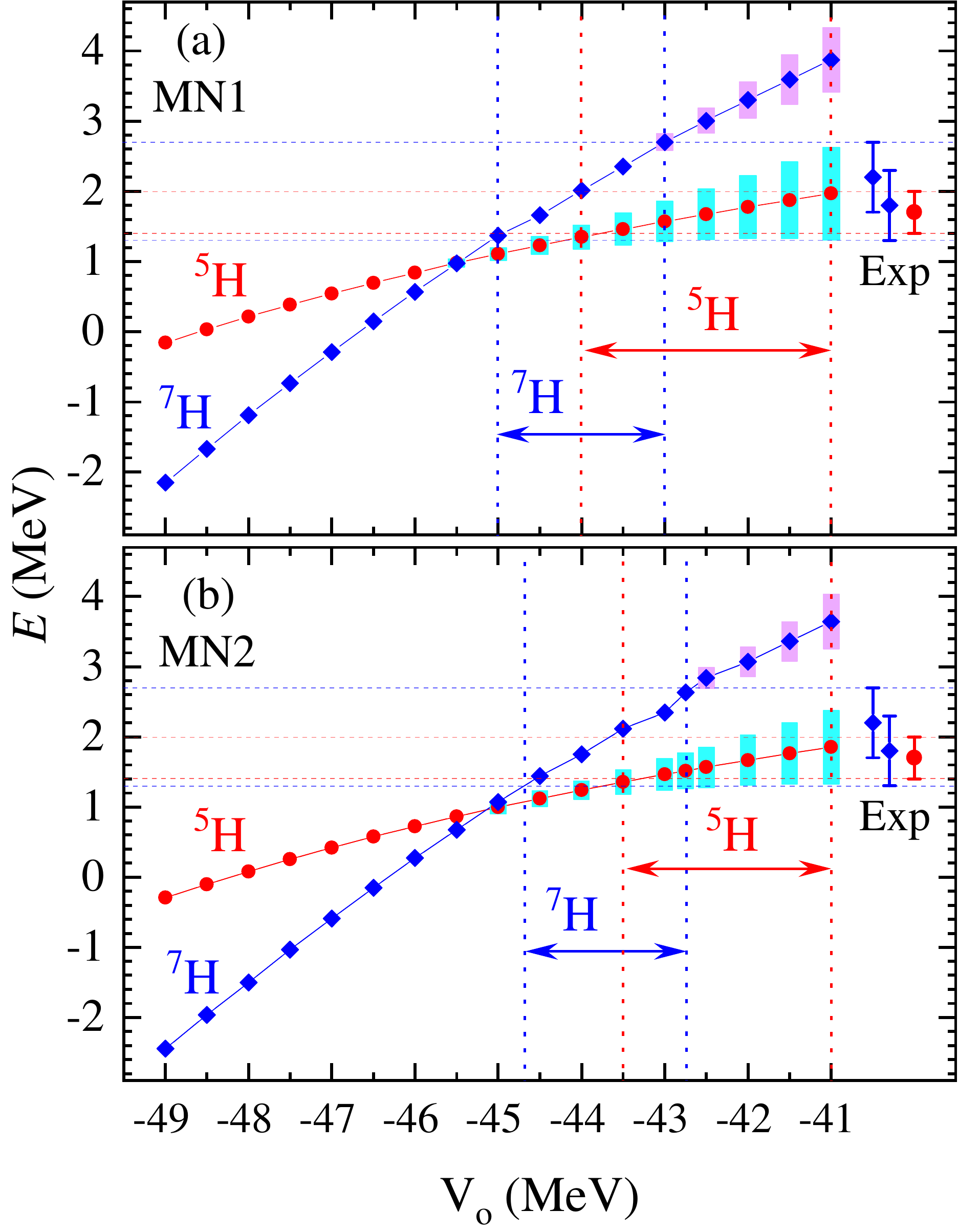}
\caption{Energies (in MeV) and widths (in keV) of $^{5,7}$H in GSM using the MN1 and MN2 interactions (see text for definition) as a function of the WS core potential depth mimicking the $^3$H core. Experimental values (Exp) are provided with error bars (see also Tab.~\ref{H_isotopes_experimental}). The minimal and maximal values of the energies of $^{5,7}$H provided by experiments are indicated by dashed lines and arrows. }{\label{H_MN1_MN2}}
\end{figure}

\begin{figure}[!htb]
\includegraphics[width=1.00\columnwidth]{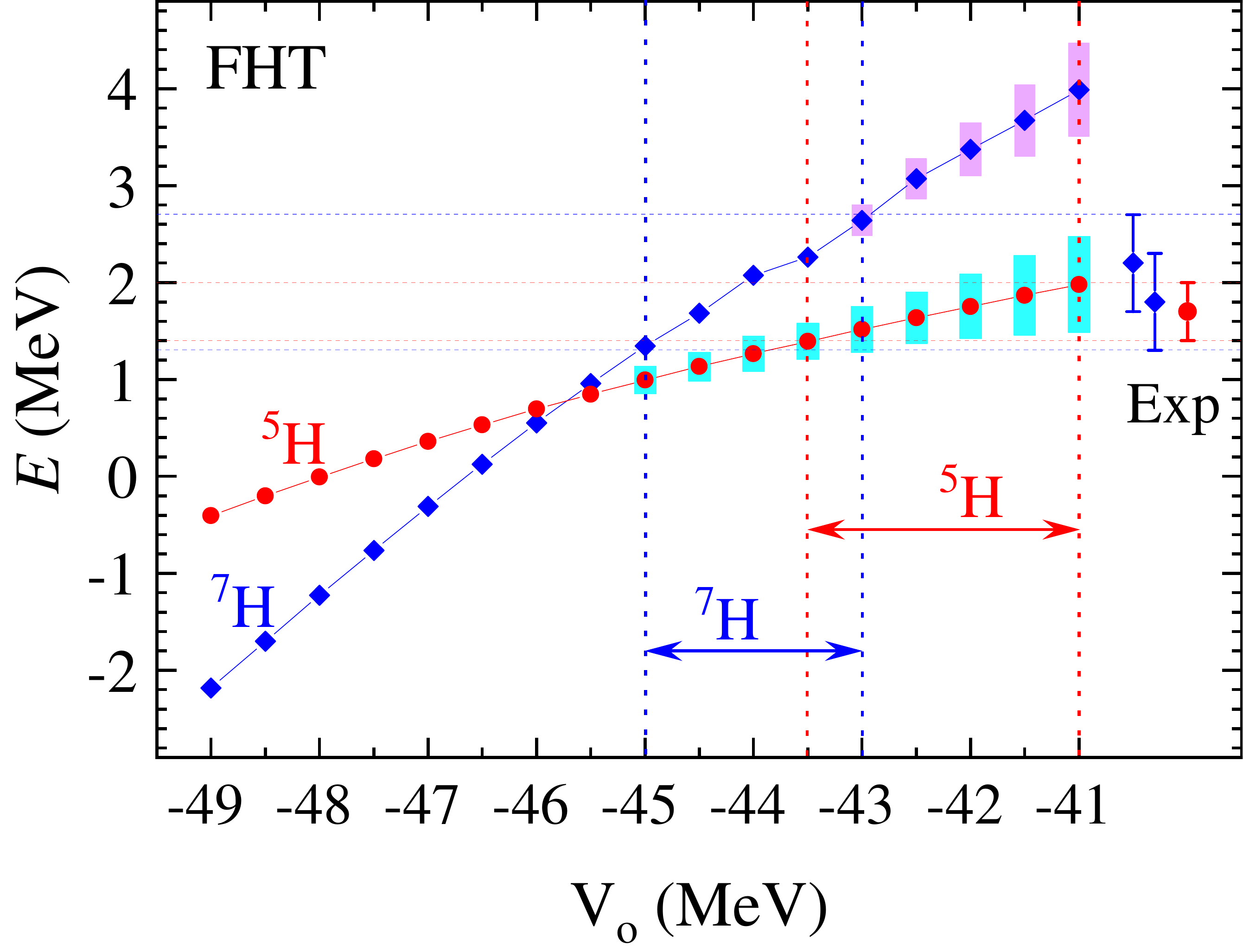}
\caption{Same as Fig.~\ref{H_MN1_MN2}, but for the FHT interaction (see text for definition).}{\label{H_FHT3}}
\end{figure}

The parameters of the WS core potential mimicking the effect of the $^3$H core remain to be considered. As energies mainly depend on its central depth, only the central potential depth will be varied, while all the other parameters are fixed. We will then change the central potential depth so that the energy of $^5$H will vary from a few hundreds of keV in negative value with respect to the $^3$H core to its highest value compatible with experimental data. Results are shown in Fig.~\ref{H_MN1_MN2} and Fig.~\ref{H_FHT3} for the MN and FHT interactions, respectively. In principle, it would be possible to calculate the theoretical errors on energies and widths by using statistical formulas similarly to that done in Ref.\cite{PhysRevC.96.054316}. However, only four hydrogen states would be considered for that matter, on the one hand, and experimental data differ much from an experiment to another, on the other hand. Thus, the theoretical errors obtained from a statistical study might not be reliable. Hence, we preferred to assess errors by considering the variations of both energies and widths inside the ranges of potential depths determined by experimental data (see Figs.~\ref{H_MN1_MN2},\ref{H_FHT3}).
Indeed, a range of core potential depths allowing for the reproduction of both the binding energies of $^5$H and $^7$H arises. Consequently, we assume that the theoretical error made on energies and widths of $^{5,7}$H is directly given by the most extreme values obtained for energy and width in this overlapping zone.

\begin{figure}[!htb]
\includegraphics[width=1.00\columnwidth]{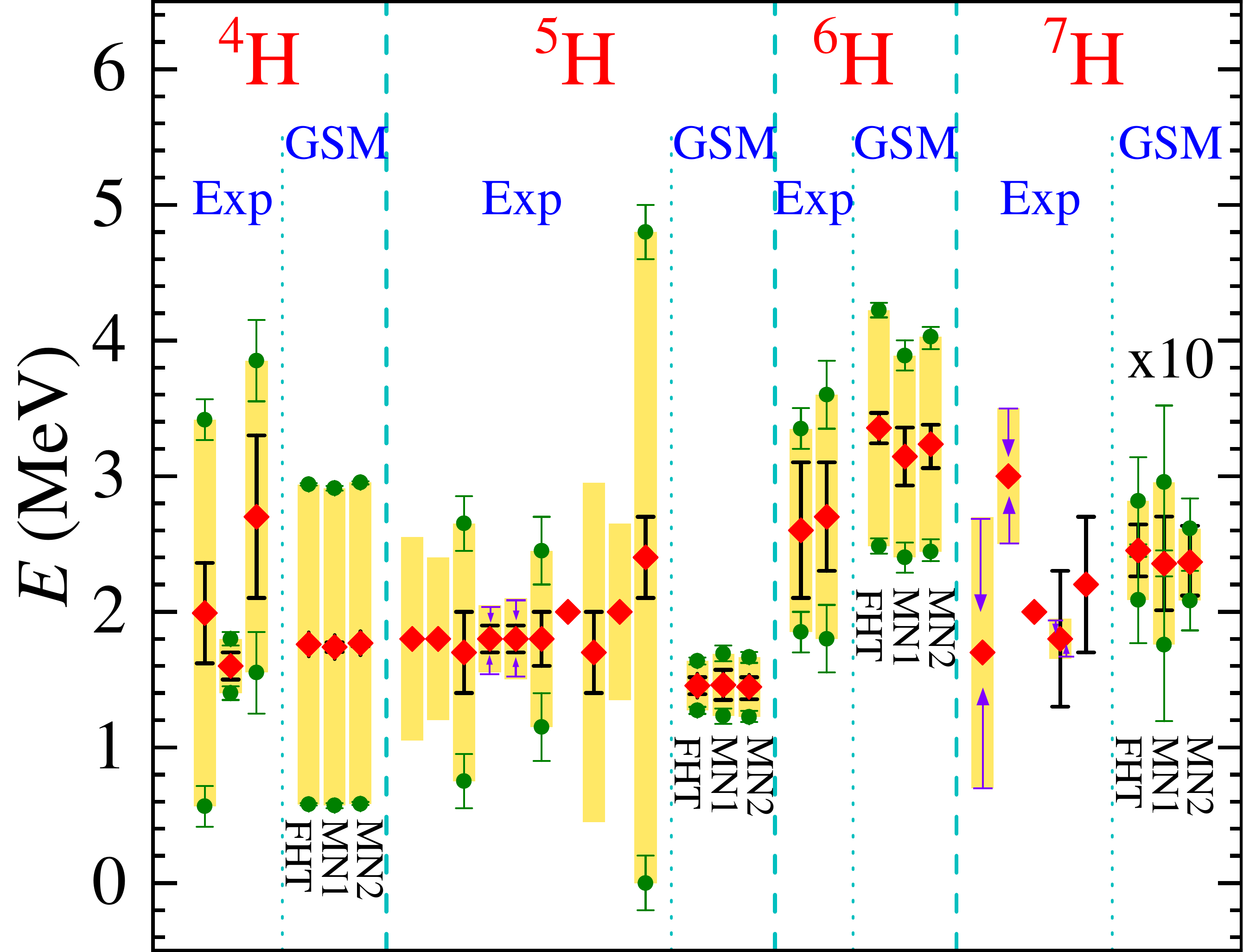}
\caption{Energies and widths (in MeV) of the hydrogen isotopes obtained from GSM calculations using the FHT, MN1 and MN2 as two-body interaction (see text for definition) and compared to experimental values. The theoretical error bars are determined from the difference between the minimal and maximal energies and widths obtained using different core WS potentials (see text for details). Experimental data, enumerated in Tab.~\ref{H_isotopes_experimental}, are either given along with error bars or maximal values for the case of two experiments related to $^7$H. Theoretical widths of $^7$H and associated errors have been enlarged by a factor 10 for readability.}{\label{H_iso_exp2}}
\end{figure}

We can then now consider all the $^{4-7}$H isotopes with the devised Hamiltonians, defined by both the two-body interaction, of MN1, MN2 or FHT type, and the one-body WS potential of the $^3$H core, whose potential depth varies as stipulated above. The different experimental energies and widths of considered hydrogen isotopes obtained from several experiments have been summarized in Tab.~\ref{H_isotopes_experimental}.  Our calculated results are depicted in Fig.~\ref{H_iso_exp2} and all energies are given with respect to the $^3$H core. One can see that theoretical calculations are in good agreement with experimental data. Indeed, both energies and widths are compatible with experimental values. Note that the width of $^4$H is larger than that obtained in Ref.\cite{PhysRevC.104.024319} by a factor 2. However, this can be explained by the fact that the structure of $^4$H is hereby oversimplified, as it is one neutron above the $^3$H core, and hence is only sensitive to the one-body WS potential of the Hamiltonian. Added to that, width varies quickly with energy, so that a difference in energy of a few hundreds of keV can easily provide 1-2 MeV difference for width. Note that the width is still comparable to experimental data, despite the one-body nature of $^4$H in our model. 
The energy and width of $^5$H are rather well constrained in our calculations, as the estimated theoretical errors for both energy and width are about 100 keV and 200 keV when using the FHT and MN interactions, respectively. $^5$H is then predicted to be a resonance state of about 1.4 MeV for energy and 500 keV for width. One can note that our results are very different from those of Refs.\cite{PhysRevC.62.014312,PhysRevC.62.014312,Lazauskas2019}, where $^5$H is predicted to be a broad resonance of width around 2.5-3.5 MeV. Nevertheless, this result is compatible with those of Refs.\cite{PhysRevC.63.027001,ADAHCHOUR2008252}, where a width around 1 MeV is predicted for $^5$H. The results for $^6$H do not vary much according the used Hamiltonian either, as $^6$H has an energy of about 3.2 MeV and a width of about 2 MeV over all the whole range of considered potential depths and with both the MN and FHT interactions. Conversely, the energy of $^7$H has an estimated error of about 400 keV and 600 keV when using the FHT and MN interactions, respectively. This is due to both the larger experimental error on the energy of $^7$H and the faster variations of the theoretical energy of $^7$H with respect to the potential strength of the one-body WS potential of the $^3$H core. The neutron-emission width of the ground state of $^7$H exhibits variations similar to that of $^5$H, as its value lies between 15 and 240 keV with the MN interaction and between 10 and 140 keV with the FHT interaction (see Figs.~\ref{H_MN1_MN2},\ref{H_FHT3}). Even though one has rather large error bars for the energy and width of $^7$H, this isotope remains a narrow resonance in all situations.
The closure of the $0p_{3/2}$ neutron shell can be reasonably seen responsible for the small width of the $^7$H ground state. Indeed, it is this subshell closure which generates the abnormally large binding energy of $^8$He \cite{PhysRevC.67.054311,PhysRevC.96.054316}. Note that, similarly to Refs.\cite{PhysRevC.104.024319,PhysRevLett.119.032501,PhysRevC.100.054313}, our calculations provide the width values compatible with the smallest experimental widths obtained from the various experiments dedicated to the $^{5-7}$H isotopes.\\

\textit{Summary.}~
Hydrogen isotopes remain among the least known nuclei of the beginning of the nuclear chart. Even though hydrogen isotopes possess fewer than 10 nucleons, their theoretical study is made difficult due to their broad resonance character when $A \geq 4$. Consequently, only the well bound deuteron and triton ground states are well known.

Hence, for the first time, up to our knowledge, a direct calculation of the energy and width of the ground states of $^{6,7}$H has been done in this paper in the frame of the complex-energy shell model. In order to obtain results without recurring to model space truncation, one devised a core+valence neutron particles using a $^3$H core. In order to minimize dependence on theoretical assumptions, the two-body residual interaction has been taken from an independent calculation of helium isotopes. Added to that, the WS potential arising from the $^3$H core was allowed to vary in a physical range determined from available experimental data. The latter procedure also allowed to assess the theoretical errors present in the calculated values of energy and width of $^{4-7}$H. 

Our calculations provided with results compatible with current experimental data (see Tab.~\ref{H_isotopes_experimental} and Figs.~\ref{H_MN1_MN2},\ref{H_FHT3},\ref{H_iso_exp2}). Even though the estimated theoretical errors are significant for $^7$H, our calculations showed that $^7$H is expected to be a very narrow resonance. The small width of $^7$H seems to have the same origin as the abnormally large binding energy of $^8$He, i.e.~it might arise from the closure of the $0p_{3/2}$ neutron shell. The calculated widths of the $^{5,7}$H isotopes compare well with the smallest widths arising from experimental studies (see Tab.~\ref{H_isotopes_experimental} and Figs.~\ref{H_MN1_MN2},\ref{H_FHT3},\ref{H_iso_exp2}). Hence, we strongly suggest experimentalists to focus on the hydrogen isotopes beyond neutron-drip line, as it is very likely that $^{5,7}$H belong to the most narrow neutron resonances of the beginning of the nuclear chart. If this assumption is confirmed, they would then be the most asymmetric many-body narrow resonances with $^{10}$He, and could provide for sure important new information about the nucleon-nucleon interaction under extreme conditions.\\

\textit{Acknowledgments.}~
We thank Zaihong Yang for his suggestions and useful comments.
This work has been supported by the National Natural Science Foundation of China under Grant Nos. 11921006, 12175281 and 11975282; the Strategic Priority Research Program of Chinese Academy of Sciences under Grant No. XDB34000000; the Key Research Program of the Chinese Academy of Sciences under Grant No. XDPB15; the State Key Laboratory of Nuclear Physics and Technology, Peking University under Grant No. NPT2020KFY13.

\bibliography{Ref}
\end{document}